\documentclass[11pt]{ws-procs961x669}
\usepackage{geometry}
\usepackage{tabularx,adjustbox}
\usepackage[retainorgcmds]{IEEEtrantools}
\usepackage[all,knot,poly]{xy}

\usepackage[utf8]{inputenc}
\usepackage{multirow}
\usepackage{array}

\usepackage{mathtools}
\usepackage{etoolbox}
\usepackage{makecell}
\usepackage{diagbox}
\usepackage{soul}
\usepackage[nodisplayskipstretch]{setspace}
\usepackage{rotating}
\usepackage{color}
\numberwithin{equation}{section}
\usepackage{multicol}
\usepackage{empheq}
\usepackage{enumitem}
\usepackage{verbatim}
\usepackage{microtype}
\usepackage[super,sort&compress]{natbib} 

\makeatletter
\renewcommand{\ps@plain}{%
\renewcommand{\@oddhead}{\hfil{\footnotesize%
A contribution to the Julian Schwinger Centennial Conference, %
7--12 February 2018, Singapore}\hfil}%
\renewcommand{\@evenhead}{\@oddhead}%
\renewcommand{\@oddfoot}{\hfil\thepage}%
\renewcommand{\@evenfoot}{\thepage\hfil}%
}
\makeatother
\crop[off]

\begin{document}
\singlespacing

\vspace{-3.3em}
\title{\Large Are Dyons the Preons of the Knot Model?}
\author{Robert J. Finkelstein}
\address{Physics \& Astronomy, University of California, Los Angeles, \\
475 Portola Plaza, Los Angeles, California 90095, USA \\
finkel@physics.ucla.edu}

\begin{abstract}
We consider the possibility that the preons defined by the SLq(2) extension of the Standard Model may be identified with Schwinger dyons. The SLq(2) extension is here presented as a model that may exist in either a currently observable electric phase or in a magnetic phase that is predicted but currently unobservable. 

\end{abstract}
\bodymatter
\section{Introduction}

\linespread{1.45}\selectfont

At this centennial for Julian, I would like to repeat the words that he spoke at the memorials for Tomonaga and George Green. As he said of them and now is true of himself, 
"Julian lives on in the minds and hearts of the many people whose lives he touched and graced. And he is, in a manner of speaking, alive, well, and living among us."
I would like to pursue one of the seminal thoughts that he has left with us.

In 1948 Dirac\cite{dirac} attempted to widen Maxwell theory by the introduction of magnetic poles. This idea was further developed in 1969 with Schwinger's paper\cite{schwinger} entitled ``A Magnetic Model of Matter'' where it was suggested that the strong nuclear coupling stemmed from the Dirac magnetic field, and it was further proposed that the most elementary particles, which he named ``dyons,'' carried both electric and magnetic charge. Since any preon is presumably smaller and heavier than the leptons and quarks, any preon picture suggests a very strong binding force --- which presents a theoretical challenge that has not yet been met.

There were, however, three phenomenological papers, namely: Harari\cite{harari79} (1979), Shupe\cite{shupe79} (1979), and Raitio\cite{raitio} (1980), that successfully represented the empirical data on leptons and quarks in terms of a simple preon model.

Beginning in 2005, in total ignorance of the these phenomenological papers, I began to study the possibility of extending the standard model of elementary particles, in admitting topological degrees of freedom for the field particles by replacing the field operators $\Psi(x)$ by
\begin{equation*}
    {\Psi}(x) \rightarrow \tilde{\Psi}^j_{m,m'} (x) D^j_{m,m'}(q) \tag{1.0}
\end{equation*}
where $ D^j_{m,m'}(q)$ is an irreducible representation of the knot algebra SLq(2), while $\tilde{\Psi}^j_{m,m'} (x)$ satisfies the Lagrangian of the standard model after its modification by the form factors generated by the adjoined $D^j_{m,m'}(q)$ factors \cite{finkelstein09,finkelstein14b,finkelstein15,finkelstein14a}. Unexpectedly, this topological model which we shall call the "knot model" agrees with the phenomenological models of Harari, Shupe, and Raitio.

In this extension of the standard model, the states $(j, m, m')_q$ of the SLq(2) algebra are postulated to be restricted by the topological conditions
\begin{equation}
(j,m,m')_q = \frac{1}{2}(N, w, r+o) \label{knotrestriction}
\end{equation}
where $(j,m,m')_q$ labels a state of the quantum knot and $(N,w,r)$ labels the 2d projection of a corresponding oriented classical knot. In this correspondence each quantum state $(j,m,m')_q$ is labelled by a classical knot $(N,w,r)$ so that the quantum kinematics is restricted by the spectrum of a classical knot. Since this restriction is on the states of the SLq(2) algebra and not on states of the standard model, it limits only the new degrees of freedom and does not disturb the preexisting symmetries of the Standard Model. 

In \eqref{knotrestriction} $N$, $w$, and $r$ are respectively the number of crossings, the writhe, and the rotation of the 2d projection of the corresponding classical knot. Here $o$ is an odd number required by an otherwise unacceptable difference in parity between the two sides of \eqref{knotrestriction}. We set $o=1$ for the simplest knot, the trefoil. Eqn.~\eqref{knotrestriction} then describes a correspondence between a state of the quantum knot $(j,m,m')_q$ and a 2d-projected classical knot $(N,w,r)$. The dynamical evolution of the field is still described by quantum field theory but the quantum dynamics is \textit{kinematically} constrained by classical knot topology.

If the four elements $(a,b,c,d)$ of the fundamental representation of SLq(2) are assumed to be creation operators for fermionic preons, then \emph{the creation operators for the simplest composite preonic structures that are topologically stable,} are \emph{the four quantum trefoils}: $D^{j/2}_{\frac{w}{2} \frac{r+1}{2}}$ where $j = 3$, $w = \pm 3$ and $r = \pm 2$.
It then turns out that these are the creation operators for the four families of elementary fermions as follows:
charged leptons, $D^{3/2}_{\frac{3}{2} \frac{3}{2}}$; neutrinos, $D^{3/2}_{-\frac{3}{2} \frac{3}{2}}$; down quarks, $D^{3/2}_{\frac{3}{2} -\frac{1}{2}}$; and up quarks, $D^{3/2}_{-\frac{3}{2} -\frac{1}{2}}$.
This preon representation of leptons and quarks by the knot model is in essential agreement with the preon models of Harari, Shupe, and Raitio.

There are now four independent approaches, including the present approach, based on the same empirical data, that suggest the same preonic model of leptons and quarks. These preons have, therefore, at least a virtual existence and in fact the only real question is whether they have independent degrees of freedom and can be observed, or whether they are lumps of field that concentrate mass and charge with no independent degrees of freedom, and therefore are bound.

We next consider a Schwinger dyon model in which the elementary particles, the dyons, carry both electric $(e)$ and magnetic charge $(g)$, in contrast to the models where the particles carry only electric charge.

We shall consider a SLq(2) dyon field which can exist in two phases, distinguished by two values of the deformation parameter $q$ as follows:
an $e$-phase where
\begin{equation*}
\boxed{q_e = \frac{e}{g}}
\end{equation*}
and
a $g$-phase where
\begin{equation*}
\boxed{q_g = \frac{g}{e}}
\end{equation*}
In both phases we assume $g >> e$ and therefore $q_g >> q_e$. We also assume that $e$ and $g$, and therefore $q$, are in general energy dependent, and we speculate that the dyon field may undergo transitions between the two phases over cosmological times.
It is further assumed that the elementary field particles in both phases are preons that carry both e and g charge, and that the creation operators of these dyonic preons are members of the fundamental representation of the SLq(2) algebra. To connect with observation, we study the possibility that there are composite particles in the $e$ phase, which are currently observed as leptons and quarks, while the corresponding particles of the $g$ phase are too massive to be currently produced or observed.

\section{The Two Charge Model}
We first consider a \textit{generic} field theory where the field quanta have two couplings that may be expressed in the coupling matrix
\begin{equation}
\varepsilon_q = \begin{pmatrix}
0 & \alpha_2 \\
-\alpha_1 & 0
\end{pmatrix}. \label{couplingmatrix}
\end{equation}
The two couplings $\alpha_1$ and $\alpha_2$ are assumed to be dimensionless and real and may be written as
\begin{equation}
(\alpha_1, \alpha_2) \text{ or } (\alpha_2, \alpha_1) = \left( \frac{e}{\sqrt{\hbar c}}, \frac{g}{\sqrt{\hbar c}} \right) \label{couplingconstants}
\end{equation}
where $e$ and $g$ refer to a specific two charge model and have dimensions of an electric charge.
We assume that e and g may be energy dependent and normalized at relevant energies. The reference charge is the universal constant $\sqrt{\hbar c}$.
We shall interpret the two fields presented by \eqref{couplingconstants} as describing \textit{parity conjugate} fields like the electric and magnetic fields.

The fundamental assumption that we make on this coupling matrix is that it is invariant under SLq(2) as follows
\begin{equation}
\boxed{T \varepsilon_q T^t = T^t \varepsilon_q T = \varepsilon_q} \label{slq2invariant}
\end{equation}
where $t$ means transpose and $T$ is a two dimensional representation of SLq(2):
\begin{equation}
T = \begin{pmatrix}
a & b \\
c & d
\end{pmatrix} . \label{2drepslq2}
\end{equation}
By \eqref{slq2invariant} and \eqref{2drepslq2} the elements of $T$ obey the knot algebra:
\begin{equation}
\boxed{
\begin{aligned}
ab = qba \qquad bd = qdb \qquad ad-qbc = 1 \qquad bc &= cb \\
ac = qca \qquad cd = qdc \qquad da -q_1cb = 1 \qquad q_1 &\equiv q^{-1}
\end{aligned}
}
\tag{A}
\end{equation}
\nopagebreak[4]
where
\begin{equation}
\boxed{q = \frac{\alpha_1}{\alpha_2}}
\end{equation}
so that the two couplings normalize the algebra through their ratio.

If also
\begin{equation}
\text{det} \hspace{2pt} \varepsilon_q = 1 \label{couplingdet}
\end{equation}
one has
\begin{equation}
\alpha_1 \alpha_2 = 1
\end{equation}

If the two dimensionless couplings $(\alpha_1, \alpha_2)$ are expressed in terms of $e$ and $g$, where $e$ and $g$ are the electroweak and ``gluon''-like charges, or electric and magnetic charges, then
\begin{equation}
eg = \hbar c \label{quantize}
\end{equation}
Then (2.8) implies that $q_e$ is the fine structure constant:
\begin{IEEEeqnarray}{*x+rCl+}
 &q_e & =& \frac{e}{g} \\
\text{and}& q_e&  =& \frac{e^2}{\hbar c} \sim \frac{1}{137}
\end{IEEEeqnarray}
If $g$ represents magnetic charge, then \eqref{quantize} is like the Dirac requirement according to which the magnetic charge is very much stronger than the electric charge.\cite{dirac} If the magnetic pole is very much heavier as well, it may be observable only in deep probes of space, i.e. at early and not at current cosmological temperatures, or at currently achievable accelerator energies. Since the knot form factors associated with the two phases are highly dependent on the deformation parameter $q$, the energy dependence of $e$ and $g$ will be quite different in the $e$ and $g$ phases.

We shall assume that magnetic poles do exist and shall study the possible extension of knot symmetry to magnetic charges.

The $2j+1$ dimensional representation of SLq(2), constructed on the Weyl monomial basis, may be expressed as follows
\begin{equation}
D^j_{mm'} = \sum_{n_a, n_b, n_c, n_d} A^j_{mm'}(q | n_a, n_b, n_c, n_d) a^{n_a} b^{n_b} c^{n_c} d^{n_d} \label{long}
\end{equation}
Here $a, b, c, d$ satisfy the knot algebra (A) and $n_a, n_b, n_c, n_d$ are summed over all positive integers and zero that satisfy the following equations:\cite{finkelstein14b}$^,$\cite{finkelstein15}
\begin{singlespacing} \vspace{-6mm}
\begin{IEEEeqnarray}{rCl}
n_a + n_b + n_c + n_d &=& 2j \label{n2j}\\
n_a + n_b - n_c - n_d &=& 2m \label{n2m}\\
n_a - n_b + n_c - n_d &=& 2m' \label{n2m'}
\end{IEEEeqnarray}
\vspace{-6mm}
\end{singlespacing}     

Here \cite{finkelstein14b}$^,$\cite{finkelstein15}
\begin{equation}
A^j_{mm'} (q \vert n_a n_b n_c n_d) = \left [ \frac{ \langle n_+ ' \rangle_1 ! \langle n_- ' \rangle_1 ! }{\langle n_+ \rangle_1 ! \langle n_- \rangle_1 !} \right ]^{\frac{1}{2}} \frac{ \langle n_+ \rangle_1 !}{\langle n_a \rangle_1 ! \langle n_b \rangle_1 !} \frac{\langle n_- \rangle_1 !}{\langle n_c \rangle_1 ! \langle n_d \rangle_1 !}
\end{equation}
where $n_{\pm} = j \pm m$, $n'_{\pm} = j \pm m'$, and $<n>_q = 1 + q +... +q^{n-1}$, with $<n>_1=<n>_{q_1}$ where $q_1=q^{-1}$.

The two dimensional representation, $T$, already introduced, now reappears as the $j = \frac{1}{2}$ fundamental representation of SLq(2),
\begin{align}
D^{\frac{1}{2}}_{mm'} &= \begin{pmatrix} a & b \\ c & d \end{pmatrix} \label{jhalfrep} \\
&=T \nonumber
\end{align}
In a physical model with the $\varepsilon_q$ coupling we interpret $(a,b,c,d)$ in \eqref{long} as creation operators for $(a,b,c,d)$ particles, which we have termed preons. Then $D^j_{mm'}(a,b,c,d)$ as given by \eqref{long} is the creation operator for the quantum state $(j,m,m')_q$ containing $(n_a, n_b, n_c, n_d)$ preons.

\section{Noether Charges carried by $D^j_{mm'}$ knots\cite{finkelstein14b}}
The knot algebra (A) is invariant under
\begin{eqnarray}
U_a (1) \times U_b (1): & a' = e^{i \varphi_a} a  &\qquad b' = e^{i \varphi_b} b\\
& d' = e^{-i \varphi_a} d &\qquad c' = e^{-i \varphi_b} c \nonumber
\end{eqnarray}
    
The transformation, $U_a (1) \times U_b (1)$, on the $(a,b,c,d)$ of SLq(2) induces on the $D^j_{mm'}$ of SLq(2) the corresponding transformation\cite{finkelstein15}
\begin{IEEEeqnarray}{rCl}
D^j_{mm'} (a,b,c,d) & \rightarrow & D^j_{mm'} (a',b',c',d') \\
& = & e^{i(\varphi_a + \varphi_b)m}e^{i(\varphi_a - \varphi_b)m'}D^j_{mm'}(a,b,c,d) \\
&= & U_m(1) \times U_{m'}(1)D^j_{mm'}(a,b,c,d)
\end{IEEEeqnarray}
and on the field operators as modified by the $D^j_{mm'}$
\begin{equation}
\tilde{\Psi}^j_{mm'} \rightarrow U_m(1) \times U_{m'}(1) \tilde{\Psi}^j_{mm'} \label{fieldeigen}
\end{equation}
where the modified field operators have been expressed in (1.0) as $\tilde{\Psi}^j_{mm'}(x)D^j_{mm'}(q|a,b,c,d)$.

\emph{For physical consistency any knotted field action that is allowed must be invariant under \eqref{fieldeigen} since \eqref{fieldeigen} is induced by $U_a \times U_b$ transformations that leave the defining algebra (A) unchanged. There are then Noether charges associated with $U_m$ and $U_{m'}$ that may be described as writhe and rotation charges, $Q_w$ and $Q_r$, since $m=\frac{w}{2}$ and $m'=\frac{1}{2}(r+o)$ for quantum knots.}

For quantum trefoils we have set $o=1$, and we now define their Noether charges:
\begin{empheq}[box=\fbox]{align}
Q_w &\equiv -k_wm \equiv -k_w\frac{w}{2} \label{qwdef} \\
Q_r &\equiv -k_rm' \equiv -k_r\frac{1}{2}(r+1) \label{qrdef}
\end{empheq}
where $k_w$ and $k_r$ are undetermined charges.

The generic model based on $D^{N/2}_{\frac{w}{2} \frac{r+1}{2}}$ has been worked out in some detail as a SLq(2) extension of the standard lepton-quark model at the electroweak level.\cite{finkelstein14a,finkelstein14b,finkelstein15} It is successful when formulated as a preon theory at the electroweak level. Being a new model, however, it presents some unanswered questions and in particular it does not predict whether the preons are bound or are in fact observable. Since the hypothetical preons are much smaller and heavier than the leptons and quarks, a very strong binding force is required to permit one to regard the leptons and quarks to be composed of three observable preons. The binding force could be gravitational and it could also be dyonic as suggested by Schwinger, or it could be both. To study the dyonic model one assumes that the preons are dyons.

The question that we examine here is whether there is a formulation and interpretation of the SLq(2) topological algebra such that the knot extension of the standard model can be reinterpreted and reparameterized at high energies to realistically also describe a dyonic Lagrangian of observable dyons.

To approach this question we begin to summarize the SLq(2) extension of the standard model by first restricting the states described by the field operators, $\tilde{\Psi}^j_{mm'} (x) D^j_{mm'} (q)$, to states obeying the postulated relations \eqref{knotrestriction}
\begin{equation}
\boxed{
(j,m,m')_q = \frac{1}{2}(N, w, r+o) \tag{1.1}
}
\end{equation}
and also obeying the \textit{empirically based} relations\cite{finkelstein14a}$^,$\cite{finkelstein14b}$^,$\cite{finkelstein15}
\begin{equation}
\boxed{
6 (t, -t_3, -t_0) = (N,w,r+1)
}
\end{equation}
or by (1.1)
\begin{equation}
\boxed{
(j, m, m')_q = 3(t,-t_3,-t_0)
}
\label{empirical}
\end{equation}
Here $t$ and $t_3$ refer to isotopic spin and $t_0$ refers to hypercharge. Eqn.~\eqref{empirical} holds for $j = \frac{3}{2}$ and $t = \frac{1}{2}$ as shown in Table 1. Table 1 describes an \textit{empirical correspondence between the simplest fermions} (t=$\frac{1}{2}$) \textit{and the simplest knots} (N=3), which are the classical trefoils, and so reveals an unexpected relation between the simplest fermions and the simplest knots.
\renewcommand{\arraystretch}{2}
\begin{figure}
\begin{multicols*}{2}
\begin{singlespace}
\begin{tabular}{r c r r r | l c r r c r}
\multicolumn{10}{c}{\textbf{Table 1:} Empirical Support for $(N,w,r+1) = 6 (t, -t_3, -t_0)$} \\ 
\hline \hline
 & Elementary Particles & $t$ & $t_3$ & $t_0$ & Classical Trefoil & $N$ & $w$ & $r$ & $r+1$ & $D^{N/2}_{\frac{w}{2}\frac{r+1}{2}}$ \\[0.2cm]
\hline
\multirow{2}{*}{\hspace{-4pt}leptons $\Bigg \{$} & $(e, \mu, \tau)_L$ & $\frac{1}{2}$ & $-\frac{1}{2}$ & $-\frac{1}{2}$ & & 3 & 3 & 2 & 3  & $D^{3/2}_{\frac{3}{2}\frac{3}{2}}$\\
& $(\nu_e, \nu_{\mu}, \nu_{\tau})_L$ & $\frac{1}{2}$ & $\frac{1}{2}$ & $-\frac{1}{2}$ & & 3 & $-3$ & 2 & 3 & $D^{3/2}_{-\frac{3}{2}\frac{3}{2}}$\\
\multirow{2}{*}{quarks $\Bigg \{$} & $(d, s, b)_L$ & $\frac{1}{2}$ & $-\frac{1}{2}$ & $\frac{1}{6}$ & & 3 & 3 & $-2$ & $-1$ & $D^{3/2}_{\frac{3}{2}-\frac{1}{2}}$\\
& $(u, c, t)_L$ & $\frac{1}{2}$ & $\frac{1}{2}$ & $\frac{1}{6}$ & & 3 & $-3$ & $-2$ & $-1$ & $D^{3/2}_{-\frac{3}{2}-\frac{1}{2}}$\\ [0.2cm]
\hline
\end{tabular}
\end{singlespace}
\break
\end{multicols*} 
\caption*{\textit{The symbols $(\quad)_L$ designate the left chiral states. The topological labels $(N, w, r)$ on the right provide a way to label the left chiral states. The last column describes the states of the Quantum Trefoil.}}
\end{figure}

\vspace{3mm}
 By Table 1: $(N,w,r+1) = 6 (t, -t_3, -t_0)$  \hfill (3.8)\\
\hspace* {1em} By postulate (1.1) and (3.8): $(j, m, m')_q = 3(t,-t_3,-t_0)$  \hfill (3.9)\\
\vspace{1mm}

The trefoil has three crossings, two values of the writhe, and after its 2D projection two values of the topological rotation. Only for the particular row-to-row correspondences shown in Table 1 do (3.8) and \eqref{empirical} hold, i.e., \emph{each of the four families of fermions labelled by $(t_3, t_0)$ is uniquely correlated with a specific $(w,r)$ classical trefoil, and therefore with a specific state $D^{3/2}_{\frac{w}{2}\frac{r+1}{2}}$ of the quantum trefoil.} 

The $t_3$ doublets of the standard model now become the writhe doublets ($w = \pm 3$) of the knot model. With this same correspondence the leptons and quarks form a knot rotation doublet ($r = \pm 2$).

We now repeat earlier work that refers explicitly to the e-phase. It is repeated here since this development is compatible with the generic two charge model and it therefore also provides a possible description of the g-phase, as well as the e-phase.

Retaining the row to row correspondence established in Table 1, it is then possible to compare in Table 2 the electroweak charges, $Q_e$, of the most elementary fermions with the total Noether charges, $Q_w + Q_r$, of the simplest quantum knots, which are the quantum trefoils. There are 4 charges $Q_e$ to fix the 2 constants $k_w$ and $k_r$.

\begin{center}
\begin{singlespacing}
\adjustbox{scale=0.85}{
\begin{tabular} {c r r r c | c l | c | c |c}
\multicolumn{10}{c}{{\textbf{Table 2:} Electric Charges of Leptons, Quarks, and Quantum Trefoils}} \\
\hline \hline
\multicolumn{5}{c |}{{Standard Model}} & \multicolumn{5}{c}{{Quantum Trefoil Model}} \\
\hline
{$(f_1, f_2, f_3)$} & {$t$} & {$t_3$} &{$t_0$} & {$Q_e$} & {$(N,w,r)$} & {$D^{N/2}_{\frac{w}{2}\frac{r+1}{2}}$} & {$Q_w$} & {$Q_r$} & {$Q_w + Q_r$} \\ [0.2cm]
\hline
$(e, \mu, \tau)_L$ & $\frac{1}{2}$ & $-\frac{1}{2}$ & $-\frac{1}{2}$ & $-e$ & $(3,3,2)$ & $D^{3/2}_{\frac{3}{2} \frac{3}{2}}$ & $-k_w \left( \frac{3}{2} \right)$ & $-k_r \left( \frac{3}{2} \right)$ & $-\frac{3}{2}(k_r + k_w)$ \\
$(\nu_e, \nu_{\mu}, \nu_{\tau})_L \hspace{-10pt}$ & $\frac{1}{2}$ & $\frac{1}{2}$ & $-\frac{1}{2}$ & $0$ & $(3,-3,2)$ & $D^{3/2}_{-\frac{3}{2} \frac{3}{2}}$ & $-k_w \left( -\frac{3}{2} \right)$ & $-k_r \left( \frac{3}{2} \right)$ & $\frac{3}{2}(k_w - k_r) $ \\
$(d,s,b)_L$ & $\frac{1}{2}$ & $-\frac{1}{2}$ & $\frac{1}{6}$ & $-\frac{1}{3}e$ & $(3,3,-2)$ & $D^{3/2}_{\frac{3}{2} -\frac{1}{2}}$ & $-k_w \left( \frac{3}{2} \right)$ & $-k_r \left( -\frac{1}{2} \right)$ & $\frac{1}{2}(k_r - 3k_w)$ \\
$(u,c,t)_L$ & $\frac{1}{2}$ & $\frac{1}{2}$ & $\frac{1}{6}$ & $\frac{2}{3}e$ & $(3,-3, -2)$ & $D^{3/2}_{-\frac{3}{2} -\frac{1}{2}} \hspace{-5pt}$ & $-k_w \left( -\frac{3}{2} \right)$ & $-k_r \left( -\frac{1}{2} \right)$ & $\frac{1}{2}(k_r + 3k_w)$ \\
& & \multicolumn{3}{c |}{$ \hspace{-8pt} Q_e = e(t_3+t_0) \hspace{-10pt}$} & \multicolumn{2}{c}{\normalsize $(j, m, m'_q) = \frac{1}{2}(N, w, r + 1)$} & $Q_w = -k_w \frac{w}{2} \hspace{-5pt}$ & $Q_r = -k_r \frac{r+1}{2} \hspace{-4pt}$ & \hspace{2pt} by (3.6) and (3.7)\\ 
\hline
\end{tabular}
}
\end{singlespacing}
\end{center}

\emph{One sees that \fbox{$Q_w + Q_r = Q_e$} is satisfied for charged leptons, neutrinos and for both up and down quarks with a single value of $k$:}
\begin{equation}
k = k_r = k_w =\frac{e}{3} \label{3k}
\end{equation}
and also that $t_3$ and $t_0$ then measure the writhe and rotation charges respectively:
\begin{equation}
Q_w = et_3 \label{qwet3}
\end{equation}
\begin{equation}
Q_r = e t_0 \label{qret0}
\end{equation}

Then \fbox{$Q_w + Q_r = Q_e$} provides an alternative meaning of
\begin{equation}
Q_e = e(t_3 + t_0)
\end{equation}
of the standard model.

In SLq(2) measure $Q_e = Q_w + Q_r$ is:
\begin{equation}
Q_e = -\frac{e}{3}(m+m'), \label{qeem}
\end{equation}
where
\begin{equation}
\boxed{(m, m') = \frac{1}{2}(w, r+1).}
\end{equation}
Then
\begin{equation}
\boxed{Q_e = - \frac{e}{6}(w+r+1) \label{qeewr}}
\end{equation}
for the quantum trefoils, that represent the elementary fermions.

Then the electroweak charge is a measure of the writhe $+$ rotation of the trefoil. The total electroweak charge in this way resembles the total angular momentum as a sum of two parts where the knot rotation corresponds to the orbital angular momentum and where the localized contribution of the writhe to the charge corresponds to the localized contribution of the spin of a particle to the angular momentum, i.e. the writhe and rotation correspond to the spin and orbital angular momentum, respectively. In \eqref{qeewr} $o=1$ contributes a ``ground state charge" resembling the ground state energy of the quantum oscillator. 

We may now try to extend (3.15) beyond the trefoil, where $o = 1$, to an $o$ that depends on the knot.

Then 
\begin{equation}
\boxed{Q_e = - \frac{e}{6}(w+r+o). }
\end{equation}

The total SLq(2) charge sums the signed clockwise and counterclockwise turns that any knotted energy-momentum current makes both at the crossings and in one circuit of the 2d-projected knot. In this way, the ``handedness" or chirality of the "knot particle" determines its electroweak charge, so that \textit{chirality reduces electroweak charge to a geometrical concept} similar to the way that curvature of space-time geometrizes mass and energy. This topological measure of electroweak charge, which is suggested by the leptons and quarks, goes to a deeper level than the also exact standard electroweak isotopic-spin measure that was originally suggested by the approximate equality of masses in the neutron-proton system.

As here defined, quantum knots carry the charge expressed as both $t_3 + t_0$ and $m + m'$. The conventional \textbf{$\bf{(t_3, t_0)}$ measure of charge may be based on $\bf{SU(2) \times U(1)}$ while the $\bf{(m,m')}$ measure of charge is based on SLq(2)}. These two different measures are related at the $j=\frac{3}{2}$ level by eqn.~\eqref{empirical}: $(j,m,m')_q = 3(t,-t_3,-t_0)$, where leptons and quarks are both $t = \frac{1}{2}$ isospin and $j = \frac{3}{2}$ knot particles, and preons are $j = \frac{1}{2}$ knot particles.

\section{The Fundamental Representation}

We next extend this analysis beyond $j=\frac{3}{2}$, describing leptons and quarks, and in particular to the fundamental representation $j=\frac{1}{2}$.
This extension to other states of $j$ is here intended to include as well a specialization of the generic model $(\alpha_1, \alpha_2)$ to $(e,g)/\sqrt{\hbar c}$ where $e$ is the electroweak coupling and $g$ is a hypothetical ``magnetoweak'' coupling.
We continue with the description of the electroweak phase.
As far as we now know, a magnetoweak phase may be constructed along the same lines with $k = e/3$ replaced by $k = g/3$, but then the missing experimental support of the $g$-phase, containing $g$-leptons and $g$-quarks, must be regarded as a currently unverified prediction of this formulation of the dyon model.

We continue with the extension of \eqref{qeem} from $j = \frac{3}{2}$, representing leptons and quarks, to $j = \frac{1}{2}$, representing preons,
\vspace{-2em}
\begin{singlespacing}
\begin{center}
\hspace{\stretch{2}}
\hspace{-8.8mm}
$D^{1/2}_{mm'} (q) = $
\begin{tabular}{c|cc r}
\multicolumn{2}{r}{$\frac{1}{2}$} & $-\frac{1}{2}$ \\ \cline{2-3}
$\frac{1}{2}$ & $a$ & $b$  & \hspace{5em} $Q_e=(-e/3)(m+m')$ \\
$-\frac{1}{2}$ & $c$ & $d$ \\
\end{tabular} \hspace{\stretch{2}} (3.14)
\end{center}
\end{singlespacing}
where the Noether charge in SLq(2) measure is $Q_e=(-e/3)(m+m')$.
\emph{Hence there is one charged preon, a, with charge $-\frac{e}{3}$ or $-\frac{g}{3}$ and its charge antiparticle, d, and there is one neutral preon, b, with its antiparticle, c.} 

We define the particle-antiparticle relation with respect to either electric or magnetic charge. 

If $j = \frac{1}{2}$, then $N=1$ by the postulate \eqref{knotrestriction} relating to the corresponding classical knot
\begin{equation}
(j,m,m')_q = \frac{1}{2}(N, w, r+o) \tag{1.1}
\end{equation}
The corresponding $a,b,c,d$ classical labels of the preons cannot therefore be described as knots since they have only a single crossing. The preon labels can, however, be described as \emph{2d-projections of twisted loops with $N=1$, $w= \pm 1$ and $r=0$}, where the two loops forming the twist have rotations that cancel.

Having tentatively interpreted the fundamental representation in terms of preons, labelled by twisted loops, we next consider the general representation labelled by the knot $(N, w, r)$.

\section*{Interpretation of all $D_{mm'}^j (q \vert a,b,c,d)$}
\begin{align*}
D^j_{mm'} = \sum_{n_a, n_b, n_c, n_d} A^j_{mm'} (q \vert n_a, n_b, n_c, n_d) a^{n_a} b^{n_b} c^{n_c} d^{n_d} \tag{2.11}
\end{align*}
\emph {Every $D^j_{mm'}$, as given in \eqref{long} being a polynomial in $a,b,c,d$, can be interpreted as a creation operator for a superposition of states, where each state has $n_a, n_b, n_c, n_d$ preons.}

The creation operators for the charged leptons, $D^{3/2}_{\frac{3}{2} \frac{3}{2}}$; neutrinos, $D^{3/2}_{-\frac{3}{2} \frac{3}{2}}$; down quarks, $D^{3/2}_{\frac{3}{2} -\frac{1}{2}}$; and up quarks, $D^{3/2}_{-\frac{3}{2} -\frac{1}{2}}$ are described by \eqref{long}, where the general polynomial representation of $D_{mm'}^j$ reduces to the following \textit{monomial} representations of the quantum trefoils

Quantum Trefoils

\begin{tabularx}{\textwidth}{cccc >{\raggedleft\arraybackslash}X}
$D^{3/2}_{\frac{3}{2} \frac{3}{2}} \sim a^3$ & $D^{3/2}_{-\frac{3}{2} \frac{3}{2}} \sim c^3$ & $D^{3/2}_{\frac{3}{2} -\frac{1}{2}} \sim ab^2$ & $D^{3/2}_{-\frac{3}{2} -\frac{1}{2}} \sim cd^2$ & (4.1) \\ 
\text{charged leptons} & \text{neutrinos} & \text{down quarks} & \text{up quarks}
\end{tabularx}
Here the $(j,m,m')_q$ indices are empirically determined in Tables 1 and 2. Then (4.1) implies that \emph{charged leptons and neutrinos are composed of three $a$-preons and three $c$-preons, respectively, while the down quarks are composed of one $a$- and two $b$-preons, and the up quarks are composed of one $c$- and two $d$-preons,  in agreement with the Harari, Shupe and Raitio models, and with the experimental evidence on which their models are constructed.} Note that the number of preons equals the number of crossings ($(j = \frac{N}{2} = \frac{3}{2})$ in (4.1)).

There are only four families of ``elementary fermions" differing by the two possibilities for the writhe and the two possibilities for the rotation of the projected quantum trefoil. Each of the ``elementary fermions" has 3 states of excitation, determined by the eigenstates of $\bar{D}^{3/2}_{mm'} D^{3/2}_{mm'}$, that appear in the Higgs mass term as modified by the knot form factor.\cite{finkelstein14a} The masses in the ``electric" and ``magnetic" phase are very different since $q_g >> q_e$.

The discussion up to this point identifies the Noether charge with the electroweak charge appearing in the empirical tables 1 and 2. Since the corresponding empirical support for a physical $g$ phase has not yet been seen, we are speculating that the $g$ phase has not yet been observed because it lies at a higher energy that is so far unobservable, and that the currently observable universe is an e-state of the dyon field. A higher mass of the g-phase is consistent with its higher deformation parameter, $q$, and should be observable in deep probes of space.

\section*{The Knotted Electroweak Vectors} 

We continue with the extension of $j = 3/2$, representing leptons and quarks, to $j =1/2$, representing preons, and to $j = 3$, representing knotted electroweak vectors. 

To achieve the required $U_a(1) \times U_b(1)$ invariance of the knotted Lagrangian (and to retain the associated conservation of $t_3$ and $t_0$, or equivalently of the writhe and rotation charge), it is necessary to impose topological and empirical restrictions on the knotted vector bosons by which the knotted fermions interact. \emph{For these electroweak vector fields we assume the $t=1$ of the standard model and therefore $j=3$ and $N=6$, in accord with both \eqref{knotrestriction} and \eqref{empirical}} \vspace{-0.2em} \vspace{-0.2em}
\begin{align*}
(j,m,m')_q = \frac{1}{2}(N,w,r+o) \tag{1.1} \\
(j,m,m')_q = 3(t, -t_3, -t_0) \tag{3.9}
\end{align*}
 that hold for the elementary fermion fields and \emph{that we now assume for the knotted vector fields as shown in Table 3.}
 
\begin{center} \vspace{-1.1em}
\begin{singlespacing}
\begin{tabular}{r | r r r r l}
\multicolumn{6}{c}{\textbf{Table 3:} Knotted Electroweak Vectors $(j=3)$} \\
\hline \hline
 & $Q$ & $t$ & $t_3$ & $t_0$ & $D^{3t}_{-3t_3 - 3t_0}$ \\
 \hline
 $W^+$ & $e$ & 1 & $1$ & 0 & $D^3_{-3,0} \sim c^3d^3$ \\
 $W^-$ & $-e$ & 1 & $-1$ & 0 & $D^3_{3,0} \sim a^3b^3$ \\
 $W^3$ & 0 & 1 & 0 & 0 & $D^3_{0,0} \sim f_3(bc)$ \\
 \hline
\end{tabular}
\end{singlespacing}
\end{center}

The charged $W^+_{\mu}$ and $W^-_{\mu}$ are described by six preon monomials. The neutral vector $W^3_{\mu}$ is the superposition of four states of six preons given by \vspace{-0.2em}
\begin{align}
D^3_{00} = A(0,3)b^3c^3 + A(1,2)ab^2c^2d + A(2,1)a^2bcd^2 + A(3,0)a^3d^3 \tag{4.4} \vspace{-0.6em}
\end{align}
according to the general representation of $D_{mm'}^j$ in \eqref{long}, which is reducible by the algebra (A) to a function of the neutral operator $bc$.

\section{Presentation of the Knot Model in the Preon Representation\cite{finkelstein15}}

The elements $(a,b,c,d)$ are assumed to be creation operators for both e and g preons that carry both e and g charges. The creation operator for a general state of the knot model is given by the knot representation of $D^j_{mm'}$ as a function of $(a,b,c,d)$ and $(n_a, n_b, n_c, n_d)$. This representation (2.11) implies the constraints \eqref{n2j}, \eqref{n2m}, \eqref{n2m'} on the exponents in the following way:
\begin{equation}
D^j_{mm'} = \sum_{n_a, n_b, n_c, n_d} A^{j}_{mm'} (q \vert n_a, n_b, n_c, n_d) a^{n_a} b^{n_b} c^{n_c} d^{n_d} \tag{2.11}
\end{equation}
where $(n_a, n_b, n_c, n_d)$ are summed over all positive integers and zero that satisfy the following equations (2.12)-(2.14), and where $(a,b,c,d,)$ satisfy the knot algebra (A)
\begin{IEEEeqnarray*}{+rCl+x*}
n_a+n_b+n_c+n_d & = & 2j & (2.12) \\
n_a+n_b - n_c - n_d & =  & 2m & (2.13) \\
n_a-n_b+n_c-n_d & = & 2m' . & (2.14)
\end{IEEEeqnarray*}

The two relations defining the quantum kinematics and giving physical meaning to $D^j_{mm'}$, are again the postulated \eqref{knotrestriction}: 
\begin{equation*}
(j,m,m')_q= \frac{1}{2}(N,w, r+o) \quad \text{field (flux loop) description} \tag{1.1}
\end{equation*}
and the semi-empirical \eqref{empirical} that holds at $(j,t) = (3/2, 1/2)$: 
\begin{equation*}
(j,m,m')_q = 3(t, -t_3, -t_0)_L \quad \text{particle description} \tag{3.9}
\end{equation*} 
The preceding relations (1.1) and (3.9) imply two complementary physical interpretations of the SLq(2) relations \eqref{n2j}--\eqref{n2m'}. By \eqref{knotrestriction} and (2.12)-(2.14) one has a semi-classical field description $(N, w, \tilde{r})$ of the quantum state $(j, m, m')_q$ as follows
\begin{equation}
\left \{
{\arraycolsep=1.2pt
\begin{array}{rl}
N &= n_a + n_b + n_c + n_d \\
w &= n_a + n_b - n_c - n_d  \\
\hspace*{5mm} \tilde{r} \equiv r+o &= n_a - n_b + n_c - n_d
\end{array}}
\right
.
\end{equation}
The $(N, w, r)$ are topological coordinates that may characterize either a binding field or a flux loop.

In the last line of (5.1), where $\tilde{r} \equiv r+o$ and $o$ is the parity index, $\tilde{r}$ may be termed ``the quantum rotation," and $o$ the ``zero-point rotation."

\vspace{5mm}
By \eqref{empirical} one has a particle description $(t, t_3, t_0)$ of the same quantum state $(j, m, m')_q$.
\singlespacing
\vspace{-10mm}
\begin{equation} \label{particle_description}
{\arraycolsep=1.2pt
\begin{array}{rl}
t &= \frac{1}{6} (n_a + n_b +n_c +n_d)  \\
t_3 &= -\frac{1}{6}(n_a + n_b - n_c - n_d)  \\
t_0 &= -\frac{1}{6}(n_a - n_b + n_c - n_d)
\end{array}
}
.
\end{equation}

In \eqref{particle_description}, $(t, t_3, t_0)$ are to be read as SLq(2) preon indices agreeing with standard SU(2) $\times$ U(1) notation \textit{only at $j = \frac{3}{2}$}. In general, however, \textit{$t_3$ measures writhe charge, $t_0$ measures rotation charge and $t$ measures the total preon population or the total number of crossings of the flux-loop, in agreement with (5.1).}

\section{Interpretation of the Complementary Equations Cont.}
\vspace*{-5mm}
\singlespacing
We now present an alternative particle interpretation of the flux loop equations (5.1)
\begin{align*}
N = n_a + n_b + n_c + n_d \tag{6.1$N$} \\
w = n_a + n_b - n_c - n_d \tag{6.1$w$} \\
\tilde{r} = n_a - n_b + n_c - n_d \tag{6.1$\tilde{r}$}
\end{align*}

\emph{Equation (6.1$N$) states that the number of crossings, $N$, equals the total number of preons, $N'$, as given by the right side of this equation. Since we assume that the preons are fermions, the knot describes a fermion or a boson depending on whether the number of crossings is odd or even.} Viewed as a knot, a fermion becomes a boson when the number of crossings is changed by attaching or removing a geometric curl 
\begin{turn}{90} 
\xygraph{
    !{0;/r0.75pc/:}
    !{\xunderh}
    [u l(0.75)]!{\xbendu}
    [l (1.5)]!{\vcap[1.5]}
    !{\xbendd-}}
\end{turn}
. This picture is consistent with the view of a curl as an opened preon loop, which is in turn viewed as a twisted loop
\begin{turn}{90}
\xygraph{
    !{0;/r0.75pc/:}
    !{\xunderh}
    [u l(0.75)]!{\xbendu}
    [l (1.5)]!{\vcap[1.5]}
    !{\xbendd-}
    [l] [d(0.5)]!{\xbendu-}
    [ld]!{\vcap[-1.5]}
    [u] [r(0.5)]!{\xbendd}
}
\end{turn}
. Each counterclockwise or clockwise classical curl corresponds to a preon creation operator or antipreon creation operator respectively.

Equations (6.1$w$) and (6.1$\tilde{r}$) may also be read as particle equations as  follows.
Since $a$ and $d$ are creation operators for antiparticles with opposite charge and hypercharge, while $b$ and $c$ are neutral antiparticles with opposite values of the hypercharge, we may introduce the charged $(\nu_a)$ and neutral $(\nu_b)$ \textit{preon numbers}, similar to baryon and lepton numbers

\vspace{-10mm}
\begin{singlespacing}
\begin{align}
\nu_a &\equiv n_a -n_d \\
\nu_b &\equiv n_b - n_c
\end{align}
Then (6.1$w$) and (6.1$\tilde{r}$) may be rewritten in terms of preon numbers as
\begin{align}
&\nu_a + \nu_b = w \hspace{3pt} (= -6t_3) \\
&\nu_a - \nu_b = \tilde{r} \hspace{3pt} (=-6t_0)
\end{align}
\end{singlespacing}

\vspace{-8mm}
By (6.3) and (6.4) the conservation of the preon numbers $(\nu_a,\nu_b)$ and also of the charge and hypercharge $(t_3,t_0)$ is equivalent to the conservation of the writhe and rotation, which are topologically conserved at the 2d-classical level. In this respect, these quantum conservation laws for preon numbers correspond to the classical conservation laws for writhe and rotation.

\section{Graphical Representation of Corresponding Classical Structures}
The representation \eqref{long} of the four classical trefoils as composed of three overlapping preon loops is shown in Figure 1. In interpreting Figure 1, note that the two lobes of all the preon loops make opposite contributions to the rotation, $r$, so that the total rotation of each preon loop vanishes. When the three $a$-preons and $c$-preons are combined to form charged leptons and neutrinos, respectively, each of the three labelled circuits is counterclockwise and contributes $+1$ to the rotation while the single unlabeled and shared (overlapping) circuit is clockwise and contributes $-1$ to the rotation so that the total $r$ for both charged leptons and neutrinos is $+2$. \vspace{-0.3em}

For quarks the three labelled loops contribute $-1$ and the shared loop $+1$ so that $r=-2$.

In each case the three preons that form a lepton trefoil contribute their three negative rotation charges. The geometric and charge profile of the lepton trefoil is thus similar to the geometric and charge profile of a triatomic molecule composed of neutral atoms since the electronic valence charges of the atoms, which cancel the nuclear charges of the atoms, are shared among the atoms forming the molecule just as the negative rotation charges which cancel the positive rotation charges of the preons are shared among the preons forming the trefoils. There is a similar correspondence between quarks and antimolecules.


 

\singlespacing
\newgeometry{top=0.5cm, bottom=1.3cm, left =2cm, right=2cm}
\begin{multicols*}{2}
\begin{center}
\normalsize
\begin{tabular}{c | c}
         \multicolumn{2}{c}{\textbf{Graphical Representation of Corresponding Classical Structures}} \\ [-0.54cm]
	\multicolumn{2}{c}{\textbf{Figure 1:} Preonic Structure of Elementary Fermions} \\ 
	\begin{tabular}{c c c}
		\multicolumn{3}{r}{ \ul{$(w, r, o)$}} \\ [-0.3cm]
		\multicolumn{3}{l}{Charged Leptons, $D^{3/2}_{\frac{3}{2} \frac{3}{2}} \sim a^3$} \\ [0.2cm]
		\huge{$\varepsilon^{ijk}$} & $\hspace{17pt} a_j$ & \\ [0.4cm]
		 $\hspace{28pt} a_i$ & & $\hspace{-58pt} a_k$ \\ [-2.8cm]
		 \multicolumn{3}{l}{\xygraph{
!{0;/r1.3pc/:}
!{ \xoverh }
[u(1)] [l(0.5)]
!{\color{black} \xbendu }
[l(2)]
!{\vcap[2] =>}
!{\xbendd- \color{black}}
[d(1)] [r(0.5)]
!{\xunderh }
[d(0.5)]
!{ \xbendr}
[u(2)]
!{\hcap[2]=>}
[l(1)]
!{\xbendl- }
[l(1.5)] [d(0.25)]
!{\xbendl[0.5]}
[u(1)] [l(0.5)]
!{\xbendr[0.5] =<}
[u(1.75)] [l(2)]
!{ \xoverv}
[u(1.5)] [l(1)]
!{\color{black} \xbendr-}
[u(1)] [l(1)]
!{ \hcap[-2]=>}
[d(1)]
!{ \xbendl \color{black}}}} \\ [-1.6cm]
\multicolumn{3}{l}{ \hspace{33pt} \textcolor{red}{
\xygraph{
!{0;/r1.3pc/:}
[u(2.25)] [r(3.25)]
!{\xcapv[1] =>}
[l(2.75)] [u(.75)]
!{\xbendr[-1] =>}
[d(1)] [r(1.25)]
!{\xcaph[-1] =>}
}}}
 \\ [-0.7cm]
		 \multicolumn{3}{r}{\hspace{150pt}$(3,2,1)$}

	\end{tabular}
	&
	\begin{tabular}{c c c}
		\multicolumn{3}{r}{\ul{$(w, r, o)$}} \\ [-0.3cm]
		\multicolumn{3}{l}{$a$-preons, $D^{1/2}_{\frac{1}{2} \frac{1}{2}}$} \\ [1.58cm]
		  & &\hspace{-63pt} $a_i$ \\ [-1.3cm]
		 \multicolumn{3}{l}{\xygraph{
!{0;/r1.3pc/:}
!{\xoverv=>}
[u(0.5)] [l(1)]
!{\xbendl}
[u(2)]
!{\hcap[-2]}
!{\xbendr-}
[r(1)]
!{\xbendr}
[u(2)]
!{\hcap[2]}
[l(1)]
!{\xbendl-}}} \\ [-1.2cm]
\multicolumn{3}{l}{\hspace{39pt} \textcolor{red}{\xygraph{
!{0;/r1.3pc/:}
[d(0.75)] [l(0.75)]
!{\xcaph[-1]=>}
}
}} \\ [-0.7cm]
		 \multicolumn{3}{r}{\hspace{150pt}$(1,0,1)$}

	\end{tabular} \\ [-0.1cm] \hline
\begin{tabular}{c c c}
		\multicolumn{3}{l}{Neutrinos, $D^{3/2}_{-\frac{3}{2} \frac{3}{2}} \sim c^3$} \\ [0.2cm]
		\huge{$\varepsilon^{ijk}$} & $\hspace{17pt} c_j$ & \\ [0.4cm]
		 $\hspace{28pt} c_i$ & & $\hspace{-58pt} c_k$ \\ [-2.8cm]
		 \multicolumn{3}{l}{\xygraph{
!{0;/r1.3pc/:}
!{\xunderh }
[u(1)] [l(0.5)]
!{\color{black} \xbendu}
[l(2)]
!{ \vcap[2]=>}
!{\xbendd- \color{black}}
[d(1)] [r(0.5)]
!{\xoverh}
[d(0.5)]
!{\xbendr}
[u(2)]
!{\hcap[2]=>}
[l(1)]
!{\xbendl-}
[l(1.5)] [d(0.25)]
!{ \xbendl[0.5]}
[u(1)] [l(0.5)]
!{\xbendr[0.5]=<}
[u(1.75)] [l(2)]
!{\xunderv}
[u(1.5)] [l(1)]
!{\color{black} \xbendr-}
[u(1)] [l(1)]
!{\hcap[-2]=>}
[d(1)]
!{\xbendl \color{black}}}} \\ [-1.5cm]
\multicolumn{3}{l}{\textcolor{red} {\hspace{32pt} \xygraph{
!{0;/r1.3pc/:}
[u(2.25)] [r(3.25)]
!{\xcapv[1] =<}
[l(2.75)] [u(.75)]
!{\xbendr[-1] =<}
[d(1)] [r(1.25)]
!{\xcaph[-1] =<}
}}} \\ [-0.7cm]
		 \multicolumn{3}{r}{\hspace{145pt}$(-3,2,1)$}

	\end{tabular}
	&
	\begin{tabular}{c c c}
		\multicolumn{3}{l}{$c$-preons, $D^{1/2}_{-\frac{1}{2} \frac{1}{2}}$} \\ [1.58cm]
		  & &\hspace{-63pt} $c_i$ \\ [-1.3cm]
		 \multicolumn{3}{l}{\xygraph{
!{0;/r1.3pc/:}
!{\xunderv=>}
[u(0.5)] [l(1)]
!{\xbendl}
[u(2)]
!{\hcap[-2]}
!{\xbendr-}
[r(1)]
!{\xbendr}
[u(2)]
!{\hcap[2]}
[l(1)]
!{\xbendl-}}} \\ [-1.2cm]
\multicolumn{3}{l}{\textcolor{red}{\hspace{39pt} \xygraph{
!{0;/r1.3pc/:}
[d(0.75)] [l(0.75)]
!{\xcaph[-1]=<}
}}} \\ [-0.7cm]
		 \multicolumn{3}{r}{\hspace{145pt}$(-1,0,1)$}

	\end{tabular} \\ [-0.1cm] \hline
\begin{tabular}{c c c}
		\multicolumn{3}{l}{$d$-quarks, $D^{3/2}_{\frac{3}{2} -\frac{1}{2}} \sim a_ib^2$} \\ [0.2cm]
		\huge{$\varepsilon^{ijk}$} & $\hspace{19pt} b_i$ & \\ [0.4cm]
		 $\hspace{28pt} a_i$ & & $\hspace{-56pt} b_k$ \\ [-2.8cm]
		 \multicolumn{3}{l}{\xygraph{
!{0;/r1.3pc/:}
!{\xoverh }
[u(1)] [l(0.5)]
!{\color{black} \xbendu}
[l(2)]
!{\vcap[2]=<}
!{\xbendd- \color{black}}
[d(1)] [r(0.5)]
!{ \xoverv}
[u(0.5)] [r(1)]
!{ \xbendr }
[u(2)]
!{\hcap[2]=<}
[l(1)]
!{\xbendl- }
[u(0.5)] [l(3)]
!{\xoverv}
[u(1.5)] [l(1)]
!{\color{black} \xbendr- }
[l(1)] [u(1)]
!{\hcap[-2]=<}
[d(1)]
!{\xbendl  \color{black}}
[r(2)] [u(0.5)]
!{\xcaph- =>}}} \\[-1.6cm]
\multicolumn{3}{l}{\hspace{42pt} \textcolor{red}{\xygraph{
!{0;/r1.3pc/:}
[u(1.75)] [l(1.5)]
!{\xbendd[-1]=<}
[u(0.75)] [r(1)]
!{\xbendl[-1]=<}
[d(1)] [l(2.25)]
!{\xcaph[-1]=>}
}}} \\ [-0.7cm]
		 \multicolumn{3}{r}{\hspace{140pt}$(3,-2,1)$}

	\end{tabular}
	&
	\begin{tabular}{c c c}
		\multicolumn{3}{l}{$b$-preons, $D^{1/2}_{\frac{1}{2} -\frac{1}{2}}$} \\ [1.58cm]
		  & &\hspace{-63pt} $b_i$ \\ [-1.3cm]
		 \multicolumn{3}{l}{\xygraph{
!{0;/r1.3pc/:}
!{\xoverv=<}
[u(0.5)] [l(1)]
!{\xbendl}
[u(2)]
!{\hcap[-2]}
!{\xbendr-}
[r(1)]
!{\xbendr}
[u(2)]
!{\hcap[2]}
[l(1)]
!{\xbendl-}}} \\ [-1.5cm]
\multicolumn{3}{l}{\hspace{40pt} \textcolor{red}{\xygraph{
!{0;/r1.3pc/:}
[u(0.75)] [l(0.75)]
!{\xcaph[1]=>}
}}} \\ [-0.1cm]
		 \multicolumn{3}{r}{\hspace{145pt}$(1,0,-1)$}

	\end{tabular} \\ [-0.1cm] \hline
\begin{tabular}{c c c}
		\multicolumn{3}{l}{$u$-quarks, $D^{3/2}_{-\frac{3}{2} -\frac{1}{2}} \sim c_id^2$} \\ [0.2cm]
		 \huge{$\varepsilon^{ijk}$} & $\hspace{19pt} d_i$ & \\ [0.4cm]
		 $\hspace{28pt} c_i$ & & $\hspace{-56pt} d_k$ \\ [-2.8cm]
		 \multicolumn{3}{l}{\xygraph{
!{0;/r1.3pc/:}
!{\xunderh }
[u(1)] [l(0.5)]
!{\color{black} \xbendu}
[l(2)]
!{\vcap[2] =<}
!{\xbendd- \color{black}}
[d(1)] [r(0.5)]
!{\xunderv}
[u(0.5)] [r(1)]
!{\xbendr}
[u(2)]
!{\hcap[2] =<}
[l(1)]
!{\xbendl-}
[u(0.5)] [l(3)]
!{\xunderv}
[u(1.5)] [l(1)]
!{\color{black} \xbendr-}
[l(1)] [u(1)]
!{\hcap[-2]=<}
[d(1)]
!{\xbendl \color{black}}
[r(2)] [u(0.5)]
!{\xcaph- =>}}} \\[-1.75cm]
\multicolumn{3}{l}{\hspace{44pt}\textcolor{red}{\xygraph{
!{0;/r1.3pc/:}
[u(1.75)] [l(1.5)]
!{\xbendd[-1]=>}
[u(0.75)] [r(1)]
!{\xbendl[-1]=>}
[d(1)] [l(2.25)]
!{\xcaph[-1]=<}
}}} \\ [-0.7cm]
		 \multicolumn{3}{r}{\hspace{140pt}$(-3,-2,1)$}

	\end{tabular}
	&
	\begin{tabular}{c c c}
		\multicolumn{3}{l}{$d$-preons, $D^{1/2}_{-\frac{1}{2} -\frac{1}{2}}$} \\ [1.58cm]
		  & &\hspace{-63pt} $d_i$ \\ [-1.3cm]
		 \multicolumn{3}{l}{\xygraph{
!{0;/r1.3pc/:}
!{\xunderv=<}
[u(0.5)] [l(1)]
!{\xbendl}
[u(2)]
!{\hcap[-2]}
!{\xbendr-}
[r(1)]
!{\xbendr}
[u(2)]
!{\hcap[2]}
[l(1)]
!{\xbendl-}}} \\ [-1.5cm]
\multicolumn{3}{l}{\hspace{39pt} \textcolor{red}{\xygraph{
!{0;/r1.3pc/:}
[u(0.75)] [l(0.75)]
!{\xcaph[1]=<}
}}} \\ [-0.1cm]
		 \multicolumn{3}{r}{\hspace{140pt}$(-1,0,-1)$} \\
	\end{tabular} \\ 
    \multicolumn{2}{l}{\Large $Q = -\frac{e}{6}(w+r+o)$, and $(j, m, m')_q = \frac{1}{2}(N, w, r+o)$} \\ 
    \multicolumn{2}{l}{\Large $D_{mm'}^j = D^{\frac{N}{2}}_{\frac{w}{2}\frac{r+o}{2}}$} \\
    \multicolumn{2}{l}{\Large The clockwise and counterclockwise arrows are given opposite weights $(\mp 1)$} \\
    \multicolumn{2}{l}{\Large respectively. The (rotation/writhe) charge is measured by the sum of the} \\
   \multicolumn{2}{l}{\Large weighted (black/red) arrows.} \\
\end{tabular}
\end{center}
\end{multicols*}

\restoregeometry
\section{Model of Preonic Trefoil with Binding Field}

Since one may interpret the elements $(a,b,c,d)$ of the SLq(2) algebra as creation operators for either preonic particles or flux loops, the $D^j_{mp}$ may be interpreted as a creation operator for a composite particle composed of either preonic particles $(N', \nu_a, \nu_b)$ or flux loops $(N, w, \tilde{r})$ where $\nu_a$ and $\nu_b$ are charged and neutral preon numbers and $N'$ is the total number of preons. \emph{These two complementary views of the same particle may be reconciled as describing $N$-preon systems bound by a knotted field having $N$-crossings with the preons at the crossings as illustrated for $N=3$ in Figure 2.} In the limit where the three outside lobes become small or infinitesimal compared to the central circuit, the resultant structure will resemble a three particle system tied together by a string.

\section{Alternate Interpretation}

In the model suggested by Fig. 2 the parameters describing the preons and the parameters describing the flux loops may be understood as codetermined by the eigenvalues of a common Hamiltonian. On the other hand, in an alternative interpretation of complementarity, the hypothetical preons conjectured to be present in Figure 2 carry no independent degrees of freedom and may simply describe \emph{concentrations of energy and momentum at the crossings of the flux tube}. In this interpretation of complementarity, $(t, t_3, t_0)$ and $(N,w, \tilde{r})$ are just two ways of describing the same quantum trefoil of field. In this picture the preons are bound, i.e. they do not appear as free particles. This view of the elementary particles as non-singular lumps of field or as solitons has also been described as a unitary field theory\textsuperscript{10}.

\vspace{-10mm}
\begin{figure}[!b]
\begin{center}
\normalsize
\begin{tabular}{c | c}
	\multicolumn{2}{c}{\textbf{Figure 2:} Leptons and Quarks Pictured as Three Preons Bound by a Trefoil Field} \\ [-0.0cm]
	\begin{tabular}{c c c}
		\multicolumn{3}{r}{ \ul{$(w, r, o)$}} \\ [-0.3cm]
		\multicolumn{3}{l}{Neutrinos, $D^{3/2}_{-\frac{3}{2} \frac{3}{2}} \sim c^3$} \\ [0.2cm]
		 \huge{$\varepsilon^{ijk}$} & $\hspace{17pt} c_j$ & \\ [-.22 cm]
		 \multicolumn{3}{c}{\hspace{-1.655 cm}\vspace{0.22 cm}\textcolor{blue} {\scalebox{2}{\Huge .}}} \\ [-0.6cm]
		 $\hspace{28pt} c_i$ {\hspace{.33cm}\textcolor{blue} {\scalebox{2}{\Huge .}}} \hspace{-1cm} & & {\hspace{-2.65cm}\textcolor{blue} {\scalebox{2}{\Huge .}}} \hspace{2.15cm} $\hspace{-56pt} c_k$ \\ [-2.8cm]
		 \multicolumn{3}{l}{\xygraph{
!{0;/r1.3pc/:}
!{\xunderh }
[u(1)] [l(0.5)]
!{\xbendu}
[l(2)]
!{\vcap[2]=>}
!{\xbendd-}
[d(1)] [r(0.5)]
!{\xoverh}
[d(0.5)]
!{\xbendr}
[u(2)]
!{\hcap[2]=>}
[l(1)]
!{\xbendl-}
[l(1.5)] [d(0.25)]
!{\xbendl[0.5]}
[u(1)] [l(0.5)]
!{\xbendr[0.5]=<}
[u(1.75)] [l(2)]
!{\xunderv}
[u(1.5)] [l(1)]
!{\xbendr-}
[u(1)] [l(1)]
!{\hcap[-2]=>}
[d(1)]
!{\xbendl}}} \\[-1.5cm]
\multicolumn{3}{l}{\textcolor{red} {\hspace{32pt} \xygraph{
!{0;/r1.3pc/:}
[u(2.25)] [r(3.25)]
!{\xcapv[1] =<}
[l(2.75)] [u(.75)]
!{\xbendr[-1] =<}
[d(1)] [r(1.25)]
!{\xcaph[-1] =<}
}}} \\ [-0.7cm]
		 \multicolumn{3}{r}{\hspace{145pt}$(-3,2,1)$}

	\end{tabular}
&
\begin{tabular}{c c c}
		\multicolumn{3}{r}{ \ul{$(w, r, o)$}} \\ [-0.3cm]
		\multicolumn{3}{l}{Charged Leptons, $D^{3/2}_{\frac{3}{2} \frac{3}{2}} \sim a^3$} \\ [0.2cm]
		\huge{$\varepsilon^{ijk}$} & $\hspace{17pt} a_j$ & \\ [-.22 cm]
		 \multicolumn{3}{c}{\hspace{-1.58 cm}\vspace{0.22 cm}\textcolor{blue} {\scalebox{2}{\Huge .}}} \\ [-0.6cm]
		 $\hspace{28pt} a_i$ {\hspace{.33cm}\textcolor{blue} {\scalebox{2}{\Huge .}}} \hspace{-1cm}  & & {\hspace{-2.65cm}\textcolor{blue} {\scalebox{2}{\Huge .}}} \hspace{2.15cm} $\hspace{-58pt} a_k$ \\ [-2.8cm]
		 \multicolumn{3}{l}{\xygraph{
!{0;/r1.3pc/:}
!{\xoverh }
[u(1)] [l(0.5)]
!{\xbendu}
[l(2)]
!{\vcap[2]=>}
!{\xbendd-}
[d(1)] [r(0.5)]
!{\xunderh}
[d(0.5)]
!{\xbendr }
[u(2)]
!{\hcap[2]=>}
[l(1)]
!{\xbendl-}
[l(1.5)] [d(0.25)]
!{\xbendl[0.5]}
[u(1)] [l(0.5)]
!{\xbendr[0.5]=<}
[u(1.75)] [l(2)]
!{\xoverv}
[u(1.5)] [l(1)]
!{\xbendr-}
[u(1)] [l(1)]
!{\hcap[-2]=>}
[d(1)]
!{\xbendl}}} \\ [-1.6cm]
\multicolumn{3}{l}{\hspace{33pt} \textcolor{red}{
\xygraph{
!{0;/r1.3pc/:}
[u(2.25)] [r(3.25)]
!{\xcapv[1] =>}
[l(2.75)] [u(.75)]
!{\xbendr[-1] =>}
[d(1)] [r(1.25)]
!{\xcaph[-1] =>}
}}} \\ [-0.7cm]
		 \multicolumn{3}{r}{\hspace{150pt}$(3,2,1)$}

	\end{tabular} \\ [-0.1cm] \hline
	\begin{tabular}{c c c}
		\multicolumn{3}{l}{$d$-quarks, $D^{3/2}_{\frac{3}{2} -\frac{1}{2}} \sim ab^2$} \\ [0.2cm]
		\huge{$\varepsilon^{ijk}$} & $\hspace{19pt} b_j$ & \\ [-.22 cm]
		  \multicolumn{3}{c}{\hspace{-1.52 cm}\vspace{0.22 cm}\textcolor{blue} {\scalebox{2}{\Huge .}}} \\ [-0.6cm]
		 $\hspace{28pt} a_i$ {\hspace{.33cm}\textcolor{blue} {\scalebox{2}{\Huge .}}} \hspace{-1cm}  & & {\hspace{-2.65cm}\textcolor{blue} {\scalebox{2}{\Huge .}}} \hspace{2.15cm} $\hspace{-56pt} b_k$ \\ [-2.8cm]
		 \multicolumn{3}{l}{\xygraph{
!{0;/r1.3pc/:}
!{\xoverh }
[u(1)] [l(0.5)]
!{\xbendu}
[l(2)]
!{\vcap[2]=<}
!{\xbendd-}
[d(1)] [r(0.5)]
!{\xoverv}
[u(0.5)] [r(1)]
!{\xbendr}
[u(2)]
!{\hcap[2] =<}
[l(1)]
!{\xbendl-}
[u(0.5)] [l(3)]
!{\xoverv}
[u(1.5)] [l(1)]
!{\xbendr-}
[l(1)] [u(1)]
!{\hcap[-2]=<}
[d(1)]
!{\xbendl}
[r(2)] [u(0.5)]
!{\xcaph- =>}}} \\ [-1.6cm]
\multicolumn{3}{l}{\hspace{42pt} \textcolor{red}{\xygraph{
!{0;/r1.3pc/:}
[u(1.75)] [l(1.5)]
!{\xbendd[-1]=<}
[u(0.75)] [r(1)]
!{\xbendl[-1]=<}
[d(1)] [l(2.25)]
!{\xcaph[-1]=>}
}}} \\ [-0.7cm]
		 \multicolumn{3}{r}{\hspace{140pt}$(3,-2,1)$}

	\end{tabular}
	&
\begin{tabular}{c c c}
		\multicolumn{3}{l}{$u$-quarks, $D^{3/2}_{-\frac{3}{2} -\frac{1}{2}} \sim cd^2$} \\ [0.2cm]
		 \huge{$\varepsilon^{ijk}$} & $\hspace{19pt} d_j$ & \\ [-.22 cm]
		  \multicolumn{3}{c}{\hspace{-1.75 cm}\vspace{0.22 cm}\textcolor{blue} {\scalebox{2}{\Huge .}}} \\ [-0.6cm]
		 $\hspace{28pt} c_i$ {\hspace{.33cm}\textcolor{blue} {\scalebox{2}{\Huge .}}} \hspace{-1cm}  & & {\hspace{-2.9cm}\textcolor{blue} {\scalebox{2}{\Huge .}}} \hspace{2.15cm} $\hspace{-56pt} d_k$ \\ [-2.8cm]
		 \multicolumn{3}{l}{\xygraph{
!{0;/r1.3pc/:}
!{\xunderh }
[u(1)] [l(0.5)]
!{\xbendu}
[l(2)]
!{\vcap[2]=<}
!{\xbendd-}
[d(1)] [r(0.5)]
!{\xunderv}
[u(0.5)] [r(1)]
!{\xbendr}
[u(2)]
!{\hcap[2]=<}
[l(1)]
!{\xbendl-}
[u(0.5)] [l(3)]
!{\xunderv}
[u(1.5)] [l(1)]
!{\xbendr-}
[l(1)] [u(1)]
!{\hcap[-2]=<}
[d(1)]
!{\xbendl}
[r(2)] [u(0.5)]
!{\xcaph- =>}}} \\ [-1.75cm]
\multicolumn{3}{l}{\hspace{44pt}\textcolor{red}{\xygraph{
!{0;/r1.3pc/:}
[u(1.75)] [l(1.5)]
!{\xbendd[-1]=>}
[u(0.75)] [r(1)]
!{\xbendl[-1]=>}
[d(1)] [l(2.25)]
!{\xcaph[-1]=<}
}}} \\ [-0.7cm]
		 \multicolumn{3}{r}{\hspace{140pt}$(-3,-2,1)$}

	\end{tabular} \\ [-0.2cm]
\multicolumn{2}{c}{The preons conjectured to be present at the crossings are suggested by the blue dots at the crossings} \\ [-0.4cm]
\multicolumn{2}{c}{ of the lepton-quark diagrams, or at the crossings of any diagram with more crossings. These diagrams}\\ [-0.4cm]
\multicolumn{2}{l}{  may also be described as superpositions of three twisted loops.}
\end{tabular}
\end{center}
\end{figure}
\break

\emph{The physical models suggested by Fig. 2 may be further studied in the context of gravitational and ``preon binding'' with the aid of generalized preon Lagrangians similar to that given in reference 9}. The Hamiltonians of these three body systems can be parametrized by degrees of freedom characterizing both the preons ($j=\frac{1}{2}$) and the binding field ($j=1 $). The masses of the leptons and quarks ($ j=\frac{3}{2}$) can be inferred from the eigenvalues of this Hamiltonian in terms of the parameters describing these three body systems.

 There is currently no experimental guidance at these conjectured energies. These three body systems are, however, familiar in different contexts, namely

\begin{center}
H$^3$ composed of one proton and two neutrons: $PN^2$, \\
$P$ composed of one down and two up quarks: $DU^2$,\\
{$N$ composed of one up and two down quarks: $UD^2$,\\
}
\end{center}

\vspace{-.5cm}
which are similar to the preon models
\vspace{-.5cm}
\begin{center}
$U$ composed of one $c$ and two $d$ preons: $cd^2$ \\
$D$ composed of one $a$ and two $b$ preons: $a b^2$
\end{center}
where $U$ and $D$ are up and down quarks.

In order to treat the preonic three body problems in a way similar to the other three body problems one needs to include in the Lagrangian all the terms expressing the degrees of freedom generated by the knot insertions. The Higgs mass term will also acquire left and right factors coming from the adjoined knot factors. All knot insertions will be products of the $\bar{D}^j_{mm'}... D^j_{mm'}$ factors.

\section{Lower Representations}
We have so far considered the states $j = 3, \frac{3}{2}, \frac{1}{2}$ representing electroweak vectors, leptons and quarks, and preons, respectively. We finally consider the states $j=1$ and $j=0$. 

In the adjoint representation $j=1$, the particles are the vector bosons by which the $j = \frac{1}{2}$ preons interact and there are two crossings of the associated classical knot. These vectors describe the interaction of the preons, including the formation of the binding field, and are different from the $j=3$ vectors by which the leptons and quarks interact.

If $j=0$, the indices of the quantum knot are
\begin{align}
(j,m,m')_q = (0,0,0)
\end{align}
and by the rule \eqref{knotrestriction} for interpreting the knot indices on the left chiral fields
\begin{align}
\frac{1}{2}(N,w,\tilde{r}) = (j, m, m')_q &= (0, 0, 0)
\end{align}
Then the $j=0$ quantum states correspond to classical loops with no crossings $(N=0)$ just as preon states correspond to classical twisted loops with one crossing. Since $N=0$, the $j=0$ states have no preonic sources of charge and therefore no electroweak interaction.  \emph{It is possible that these }$j=0$ \emph{hypothetical quantum states are realized as electroweak non-interacting loops of field flux with} $w=0$, $\tilde{r}= r+o = 0$\emph{, and }$r = \pm1$, $o =\mp1$ \emph{ i.e. with the topological rotation }$r=\pm1$. The two states $(r, o) = (+1, -1)$ and $(-1, +1)$ are to be understood as quantum mechanically coupled.

If, as we are assuming, the leptons and quarks with $j= \frac{3}{2}$ correspond to 2d projections of knots with three crossings, and if the heavier preons with $j= \frac{1}{2}$ correspond to 2d projections of twisted loops with one crossing, then if the $j=0$ states correspond to 2d projections of simple loops, one might ask if these particles with no electroweak interactions, which are smaller and heavier than the preons, are among the candidates for ``dark matter." If these $j=0$ particles predated the $j=\frac{1}{2}$ preons, one may refer to them as ``yons" as suggested by the term ``ylem" for primordial matter.

\section{Speculations about an earlier universe and dark matter if the dyons are preons} \vspace{-0.8em}
One may speculate about an earlier universe before leptons and quarks had appeared, when there was no charge, and when energy and momentum existed only in the SLq(2) $j=0$ neutral state as simple loop currents of gravitational energy-momentum. Then the gravitational attraction would bring some pairs of opposing loops close enough to permit the transition from two $j=0$ opposing simple loops into two opposing $j=\frac{1}{2}$ twisted loops. A possible geometric scenario for the transformation of two simple loops of current (yons) with opposite rotations into two $j=\frac{1}{2}$ twisted loops of current (preons) is shown in Fig. 3. To implement this scenario one would expect to go beyond the earlier considerations of this paper. Without attempting to do this, one notes according to Fig. 3 that the fusion of two yons may result in a doublet of preons as twisted loops, which might also qualify as Higgs doublets.

\newpage    \singlespacing
\begin{center}
\normalsize
\begin{tabular}{c  c  c}
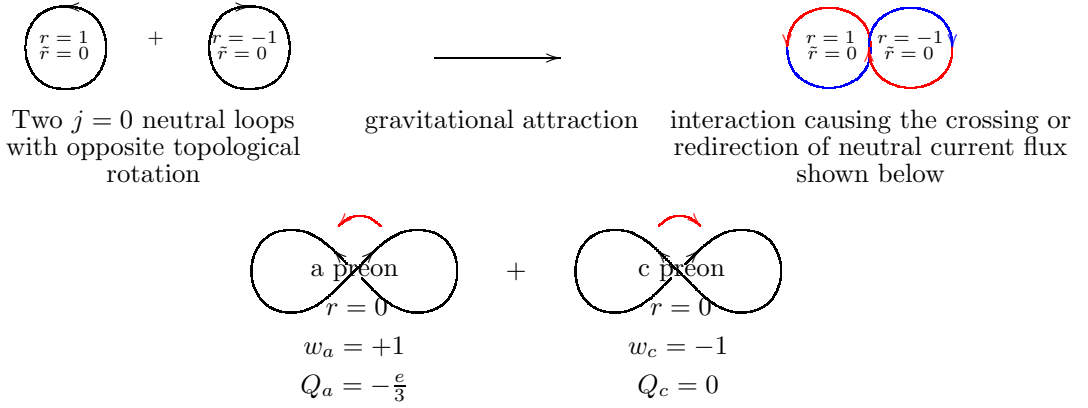
\multicolumn{3}{c}{ \textbf{Figure 3:} Creation of Preons as Twisted Loops} \\
\vspace{-3pt}
\xygraph{
!{0;/r1.3pc/:}
!{\hcap[2]=>}
!{\hcap[-2]}} \hspace{3pt}
\xygraph{
!{0;/r1.3pc/:}
!{\hcap[2]}
!{ \hcap[-2]=< }} &
\xygraph{
!{0;/r1.3pc/:}
[d(1.25)]
!{\xcaph[3]=<@(0)}} & \hspace{-4pt}
\xygraph{
!{0;/r1.3pc/:}
[d(1)]
!{\color{red} \vcap[2]=> \color{blue}}
!{\vcap[-2]}} \hspace{-7pt}
\xygraph{
!{0;/r1.3pc/:}
[d(1)]
!{\color{blue} \vcap[2]=<}
!{\color{red} \vcap[-2]=< \color{black}}}\\ [-1.7cm]
\hspace{4pt}\scriptsize{$r=1 \hspace{20pt}+ \hspace{16pt}r=-1$} & & \scriptsize{$\hspace{3pt}r=1 \hspace{8pt} r=-1$} \\ [-0.7cm]
\hspace{0pt}\scriptsize{$\tilde{r}=0 \hspace{32pt} \hspace{16pt}\tilde{r}=0$} & & \scriptsize{$\hspace{0pt}\tilde{r}=0 \hspace{10pt} \tilde{r}=0$} \\
Two $j=0$ neutral loops & gravitational attraction & interaction causing the crossing or  \\ [-0.55cm]
 with opposite topological & & redirection of neutral current flux \\ [-0.55cm]
  rotation & & shown below \\ [-0.55cm]
\end{tabular}
\end{center}
\begin{center}
\begin{tabular}{c c}
\textcolor{red}{\xygraph{
!{0;/r1.3pc/:}
[u(0.75)] [l(0.75)]
!{\xcaph[1]=>}
}} &
\textcolor{red}{\xygraph{
!{0;/r1.3pc/:}
[u(0.75)] [l(0.75)]
!{\xcaph[1]=<}
}} \\ [-1cm]
\xygraph{
!{0;/r1.3pc/:}
!{\xoverv=<}
[u(0.5)] [l(1)]
!{\xbendl}
[u(2)]
!{\hcap[-2]}
!{\xbendr-}
[r(1)]
!{\xbendr}
[u(2)]
!{\hcap[2]}
[l(1)]
!{\xbendl-}} &
\xygraph{
!{0;/r1.3pc/:}
!{\xunderv=<}
[u(0.5)] [l(1)]
!{\xbendl}
[u(2)]
!{\hcap[-2]}
!{\xbendr-}
[r(1)]
!{\xbendr}
[u(2)]
!{\hcap[2]}
[l(1)]
!{\xbendl-}} \\ [-1.1cm]
\multicolumn{2}{c}{+} \\ [-0.95cm]
\hspace{-95pt} & \hspace{-95pt} \\ [-0.9cm]
a preon & \hspace{-1pt} c preon \\ [-0.4cm]
\hspace{2pt}$r=0$ & \hspace{2pt}$r=0$ \\ [-0.4cm]
$w_a = +1$ & $w_c=-1$ \\ [-0.4cm]
$ Q_a = -\frac{e}{3}$ & $ Q_c = 0$  \\
\end{tabular}
\end{center}

In the scenario suggested by Figure 3 the opposing states are quantum mechanically entangled and may undergo gravitational exchange scattering.
%

The $\binom{c}{a}$ doublet of Fig. 3 is similar to the Higgs doublet which is independently required to be a SLq(2) singlet $(j=0)$ and a SU(2) charge doublet $(t=\frac{1}{2})$ by the mass term of the Lagrangian described in references 8 and 9. Since the Higgs mass contributes to the inertial mass, one should expect a fundamental connection with the gravitational field at this point.

If at an early cosmological time, only a fraction of the initial gas of the quantum loops, the yons, had been converted to preons and these in turn had led to a still smaller number of leptons and quarks, then most of the mass and energy of the universe would at the present time still reside in the dark loops while charge and current and visible mass would be confined to structures composed of leptons and quarks. \emph{In making experimental tests for particles of dark matter one would expect the SLq(2) $j=0$ dark loops to be greater in mass than the dark neutrino trefoils where $j= \frac{3}{2}$.}

\section{Summary Comments on the Magnetoweak Phase}
To express the correspondence between electroweak and magnetoweak sectors that suggests magnetic monopoles, we have assumed that the magnetoweak charges have the same topological origin as the electroweak charges so that they are also describable as quantum trefoils. There are then magnetoweak as well as electroweak charged leptons, quarks and preons. If all masses, both electroweak and magnetoweak, are fixed by the corresponding Higgs terms, then all masses are proportional to the eigenvalues of $\bar{D}^j_q (m,m') D^j_q (m,m')$ as follows\cite{finkelstein14a}
\begin{equation}
\left \langle n| \bar{D}^j_q (m,m') D^j_q (m,m') |n \right \rangle = f(q, \beta, n) \label{eigenD}
\end{equation}
where $|n \rangle$ are the eigenstates and $n$ labels the three states of excitation in a family of either the observed electroweak leptons and quarks or the predicted magnetoweak particles. Here $\beta$ is the value of $b$ on the ground state $|0 \rangle$. Since \eqref{eigenD} is a polynomial in $q$ and $\beta$ and of degree determined by $n$, the three lepton and quark masses in a family may be parametrized by $q$, $\beta$ and $n$. The masses of the electroweak and corresponding magnetoweak charged preons, leptons and quarks, could then be vastly different if the fine structure constant is close to its present value; since $q_e$ and $q_g$ would be very different
\begin{align}
\frac{q_e}{q_g} & = \left (\frac{e^2}{\hbar c} \right )^2 \nonumber \\
& \approx \left ( \frac{1}{137} \right )^2 .
\end{align}
The masses would also be highly dependent on $n$. 

Since all form factors generated by the $D_q^j (m,m')$ also depend on $q$ in a major way, the dynamics of the magnetoweak phase will accordingly differ seriously from the dynamics of the electroweak phase. The corresponding e-form-factors and masses have so far been only roughly discussed and there has not been an adequate parametrization of the dynamics of the electroweak phase in terms of the available experimental data. The corresponding experimental data is not available for the magnetoweak phase. In both cases one would be exploring the SLq(2) extension of models with either gravitational or dyonic binding.

At currently accessible accelerator energies, one may speculate that only the $e$-phase of the dyon field is observable, but the $g$-phase of this field is not, because the masses of the g-particles puts them out of reach of modern accelerators. On the other hand, it is possible that the cosmological temperatures of the early universe may be high enough to reveal a few g-particles, including magnetic monopoles. If the ambient temperature is high enough to produce g-particles, it may also be high enough to ionize the e-particles and thereby to produce e-preons.

\section*{Acknowledgements}
I thank E. Abers, C. Cadavid, J. Smit for help in the preparation of this talk. 

\end{document}